\documentclass[12pt,eqsecnum,preprint]{aastex}
\begin{document}
\title{The Origins of Causality Violations In Force-Free Simulations of Black Hole Magnetospheres}
\author{Brian Punsly}
\affil{4014 Emerald Street No.116, Torrance CA, USA 90503}
\email{brian.m.punsly@boeing.com or brian.punsly@gte.net}
\author{Donato Bini}
\affil{Instituto per Applicazioni della Matematica C.N.R., Naples,
  I-80131, Italy and International Center for Relativistic
  Astrophysics, I.C.R.A., University of Rome La Sapienza, I-00185
  Roma, Italy}
\email{donato.bini@icra.it}
\begin{abstract} Recent simulations of force-free, degenerate (ffde)
  black hole magnetospheres indicate that the fast mode radiated from
  (or near) the event horizon can modify the global potential
  difference in the poloidal direction orthogonal to the magnetic
  field, V, in a black hole magnetosphere. By contrast in MHD
  (meaning perfect magnetohydrodynamics, hereafter), a combination of Alfven and fast waves are required to alter V in a magnetosphere. This distinction is significant because it
  changes the causal boundary surface for evolving V from the event horizon
  (the inner fast critical surface in ffde) to the inner Alfven
  critical surface.  Secondly, there is a fundamental contradiction in a
  wave that alters V coming from near the horizon. The
  background fields in ffde satisfy the ``ingoing wave condition'' near
  the horizon (that arises from the requirement that all matter is
  ingoing at the event horizon), yet outgoing waves are radiated from
  this region in the simulation. The resolution of these two issues is
  important to our understanding of causality in black hole magnetospheres and ffde as a faithful
  representation of tenuous MHD near a black hole. Studying the properties of the waves in the simulations are useful tools to this end. It
  is shown that regularity of the stress-energy tensor in a freely
  falling frame requires that the outgoing (as viewed globally) waves
  near the event horizon are redshifted away and are ineffectual at
  changing V. It is also concluded that waves in massless MHD (ffde)
  are extremely inaccurate depictions of waves in a tenuous MHD plasma, near
  the event horizon, as a consequence black hole gravity. Any analysis
  based on ffde near the event horizon is seriously flawed. 
\end{abstract}
\section{Introduction} The Blandford-Znajek mechanism of extracting
  the rotational energy of a black hole (see \citet{blz77}) is one of
  the more celebrated theories in astrophysics. However, certain
  causality issues have been raised over the years (see \citet{pun01}
  and references therein). In particular, will a general set of initial
  conditions in tenuous MHD evolve to this solution? An axisymmetric
  ffde numerical simulation was presented in \citet{kom01} to study this
  question. It assumed ffde and involved a very complicated initial
  state. The system evolved by adjusting V with waves coming from near
  the event horizon. It is shown in this Letter that ffde
  grossly misrepresents the physics near the black hole and the simulations
  cannot be used to justify causality of the Blandford-Znajek mechanism.
\par The time evolution of the full MHD system of
  equations in a black hole magnetosphere is extremely
  complicated. For simplicity, cold MHD in the limit of zero plasma
  mass, ffde, is often used. In cold MHD, the causal structure of the
  magnetosphere is clear, both the fast and Alfven modes are required
  to modify the poloidal voltage drop across $\mathbf{B}$, or
  equivalently the field line rotation rate \citep{pun01,pun03,kom02}. Thus, it must be established outside of the
  Alfven point of the plasma flowing inward towards the black
  hole. In the simulation, ``the rotation of the magnetic
field lines starts near the black hole and propagates outwards in the form of a
torsional Alfven wave'' \citep{kom01}. Yet in \citet{kom02} it is noted
that in the initial state, the inner Alfven surface (an effective
one-way membrane for Alfven waves that permits traversal by
ingoing waves only) is located at the stationary limit surface which
is quite far form the horizon. Clearly, the ``torsional Alfven wave''
  must be a pure fast wave. As such the ffde simulation is not in accord
  with the time evolution of V in MHD that requires two wave
  polarizations. Likewise in ffde pulsar simulations, the star
  adjusts V by radiating Alfven waves \citep{spi03}. Komissarov acknowledges in a later paper
that the ``driving source'' for the Blandford-Znajek mechanism ``must
be located between the inner and outer Alfven surfaces of a black hole
magnetosphere and, thus, lay well outside of the
horizon''\citep{kom02}. Consequently, the evolution of V by the fast
  mode in the simulation is quite paradoxical from a causality perspective.
\par This Letter points out two physical inaccuracies of an ffde
  simulation that alters V with waves coming from near the
  horizon. First of all, any MHD wave near the event horizon requires very
  strong cross-field currents in order to propagate away from the
  horizon. This is in direct contradiction to the assumed ffde waves
  that drive the simulation from near the horizon. These nonforce-free
  inertial currents are required to move the plasma under the
  influence of powerful black hole gravity. Even if a plasma is
  extremely tenuous, black hole gravity is the dominant force near the
  black hole and it is inappropriate to approximate it away as in
  ffde. Secondly, the ability of an outgoing wave to alter V gets
  redshifted away as the horizon is approached. 
\par Many of the points of the Letter will be explored through an
  axisymmetric numerical simulation presented in \citet{kom01} that
   is a time evolution of the magnetosphere of the event horizon of a
   rotating black hole of mass, M, the space-time of which is
   described by the Kerr metric. The force-free conditions are given
   by the following relationships, written covariantly in terms of the
   Maxwell field strength tensor, $F^{\mu\nu}$, and four-current
   density, $J^{\mu}$, as well as in component form,
\begin{eqnarray}
&& F^{\mu\nu}J_{\nu}=0\;,\quad\rho_{e}\mathbf{E} +\frac{\mathbf{J}\times \mathbf{B}}{c} = 0\;.
\end{eqnarray}
The simulation is predicated on the following
set of equations,
\begin{eqnarray}
&& F^{\mu\nu}F_{\mu\nu}>0\:,\quad T^{\mu\nu}_{\quad ;\,\nu}=0\;,\quad
\ast F^{\mu\nu}_{\quad ;\,\nu}=0\;,\quad \ast F^{\mu\nu}F_{\mu\nu}=0\:,
\end{eqnarray}
where $T^{\mu\nu}$ is the stress-energy tensor of the electromagnetic
field.
\par The Letter begins with a very brief review of the
polarization properties of the two modes that exist in ffde, the
Alfven and fast mode. These modes determine the causal structure of an
ffde magnetosphere. The third section is a detailed description of the
initial state of the \citet{kom01} simulation. In section 4, the 2-D
ffde fast wave-fronts in the Kerr spacetime are calculated for the first
time in the literature. 
\section{Force-Free Discontinuities} The simulation evolves by the
propagation of step waves, force-free discontinuities. Since the field is degenerate and
magnetic by (1.2), there exists a time-like frame at each point of
space-time in which $\mathbf{E}$ vanishes \citep{kom01}. This is known
as a proper frame (not necessarily a coordinate frame). The structure
of the wave-front  simplifies in a proper frame, since there is no $\mathbf{E}$
 upstream. In the proper frame, the normal vector of the step wave is
 $\mathbf{n}$ (not necessarily a planar wave-front) and the magnetic
 field upstream (downstream) of the wave-front is designated as
 $\mathbf{b}_{u}$ ($\mathbf{b}_{d}$). The force-free constraint in
 (1.1), implies that all particles flow parallel to $\mathbf{b}$,
 otherwise the resulting $\mathbf{E}$ in the frame of the particles would
 drive large cross-field (nonforce-free) currents.
\par A wave that propagates an electromagnetic field that is
completely transverse to $\mathbf{n}$ is called the fast mode. By solving, the tangential component of Ampere's law
 and Faraday's law at the wave-front with the degeneracy condition,
 one finds that the wave propagates E=B at a velocity c with
 $\mathbf{E}$ orthogonal to the $\mathbf{n}$ - $\mathbf{b}_{u}$ plane and
 $\mathbf{B}$ in the $\mathbf{n}$ - $\mathbf{b}_{u}$ plane. 
\par Consider the oblique Alfven mode, with
 $\mathbf{n}$ at an angle $\theta$ to $\mathbf{b}_{u}$. The wave
 propagates at a velocity $c\cos{\theta}$ with variations of
 $\mathbf{B}$ and $\mathbf{n}\times\mathbf{E}$ orthogonal to the $\mathbf{n}$ - $\mathbf{b}_{u}$
 plane. There is also a normal component of $\mathbf{E}$.
\section{A Sample Simulation}
In \citet{kom01}, an ffde magnetosphere evolves from an initial state
that has only radial and azimuthal components of the magnetic field in
Boyer-Lindquist (B-L hereafter) coordinates. Symmetry is used to impose boundaries at the pole and the
equatorial plane (the equatorial plane is a
current sheet that switches off the azimuthal and poloidal magnetic
field). Thus, the problem reduces to an analysis in the upper right
quadrant of a plane. 
\par The initial state is actually very
complicated since it is adhoc. The details
will be described in two separate frames for illustrative
purposes. One is a global coordinate frame known as the
stationary frames at asymptotic infinity. These frames have a
four-velocity based on B-L time.\footnote{B-L coordinates
are denoted by $(t,r,\theta,\phi)$. The metric tensor, $g_{\mu\nu}$,
is expressed in B-L coordinates throughout the text and is parameterized by the
angular momentum per unit mass of the black hole, $a$. The following standard definitions are used:
 $\rho^{2}\equiv r^{2}+a^{2}\cos^{2}{\theta}$, $\Delta\equiv
r^{2}-2Mr+a^{2}$, the lapse function,
 $\alpha=\sqrt{\Delta\sin^{2}{\theta}/g_{\phi\phi}}$, vanishes at the horizon and $\Omega\equiv -g_{\phi t}/g_{\phi\phi}$, where $\beta^{\phi}_{_{F}}$ is the
 azimuthal three-velocity of the corotating frame of the magnetic
 field as viewed in the ZAMO frames and $\Omega$ is the angular
 velocity of the ZAMOs in the stationary frames. $B_{0}$ is a constant introduced in (7) of \citet{kom01}.} The local frame is the orthonormal
  one carried by the ZAMOs (zero angular momentum observers) which are located at fixed
B-L poloidal coordinates, $r$ and $\theta$ \citep{pun01}. The toroidal
magnetic field density, $B^{T}$, is the angular momentum flux per unit
magnetic flux in each magnetic flux tube in the steady state and the
global potential difference across a narrow tube of magnetic flux,
$\delta\Phi$, is $\Delta V$\citep{pun01}. In the initial state,
\begin{eqnarray}
&& B^{T}\equiv
\alpha\sqrt{g_{\phi\phi}}F_{r\theta}=-\frac{a\sin^{2}{\theta}B_{0}}{\rho^{2}}\;,\quad
\Delta V=-\frac{\Omega_{_{F}}}{2\pi c}\delta\Phi\;.
\end{eqnarray}
The initial state is characterized by a field line angular
velocity in the stationary frames that is zero everywhere,
$\Omega_{_{F}}=0$. 
The nonzero components of the electromagnetic field in the ZAMO frames are:
\begin{eqnarray}
&& B^{\phi}=-\frac{a}{\rho^{2}\sqrt{\Delta}}B_{0}\sin{\theta}\;,\quad B^{P}=B^{r}=\frac{B_{0}\sin{\theta}}{\rho\sqrt{g_{\phi\phi}}}\;,\\
&& E^{\perp} =-\beta^{\phi}_{_{F}}B^{P}=E^{\theta}=\frac{2Mra\sin^{2}{\theta}}{\rho^{2}\sqrt{\Delta
    g_{\phi\phi}}}B_{0}\;,\quad\beta^{\phi}_{_{F}}=\frac{\Omega_{_{F}}-\Omega}{c\alpha}\sqrt{g_{\phi\phi}}\;.
\end{eqnarray}
 The components of the four current in the initial state are
 $J^{\theta}=0$, and
\begin{eqnarray}
&&
\rho_{e}=-\frac{Mrac(r^{2}+a^{2})\sin{\theta}\cos{\theta}}{\pi\sqrt{\Delta
    g_{\phi\phi}}\rho^{6}}B_{0}\;,\quad J^{r}=J_{D}^{r}-\frac{acB_{0}\left(r^{2}+ a^{2}\right)\cos{\theta}}{2\pi\sqrt{\Delta}\rho^{5}}\;.
\end{eqnarray} In pulsar physics, $\rho_{e}$ is known as the
Goldreich-Julian charge density. The
displacement current, $\mathbf{J}_D$ and $J^{\phi}$ are complicated functions. 
\par The simulation proceeds due to the time
evolution of $T^{\mu\nu}$ in equation (1.2). The angular
momentum flux, $B^{T}$ in equation (3.1), is not a constant in the
flux tubes. The system evolves to a final state in which $B^{T}$
equals a constant in each flux tube (the Blandford-Znajek
solution). As this occurs, the field line angular velocity changes
form zero to approximately one-half of the horizon angular velocity
\citep{kom01}. Consequently by (3.3) and Gauss' law, as
the  so-called ``torsional Alfven wave'' propagates outward from the space-time
near the event horizon, it adjusts $\rho_{e}$ (the
global potential) and the poloidal current (that is required to support 
$B^{T}$ by Ampere's law). 
\section{Outgoing Force-Free Fast Waves in the Ergosphere} In this section,
we show that outgoing (as viewed globally) ffde fast waves are
redshifted away near the horizon. It is also demonstrated that ffde is an
inaccurate depiction of plasma waves near the horizon. 
\subsection{The Propagation Vector of 2-D Fast Waves in the Proper
  Frame} In this subsection, we compute the axisymmetric 2-D fast wave-fronts near the horizon. This is necessary so
that one can determine the Poynting flux of the waves in the proper
frame of the plasma. This provides a meaningful physical constraint
on the waves that is used in the next subsection to quantify the effects of
gravitational redshift. From (3.2) and (3.3), an orthonormal proper frame
is realized by a radial inward boost with velocity, $v^{r}$, relative
to the ZAMO frames. Denote the ZAMO basis vectors as
$\hat{e}_{\mu}$ and the proper frame basis vectors as $\bar{e}_{\mu}$:
\begin{mathletters}\begin{eqnarray}
&& \bar{e}_{0}=\gamma[\hat{e}_{0} +
v^{r}\hat{e}_{r}]\;,\quad\bar{e}_{r}=\gamma[v^{r}\hat{e}_{0} +
\hat{e}_{r}]\;,\quad
\bar{e}_{\theta}=\hat{e}_{\theta}\;,\quad\bar{e}_{\phi}=\hat{e}_{\phi}\;,\\
&& v^{r}=-\frac{2Mr\sin{\theta}}{\rho\sqrt{g_{\phi\phi}}}\;,\quad\gamma=\sqrt{\frac{\Delta(\rho^{2}+2Mr)+4M^{2}r^{2}}{\Delta(\rho^{2}+2Mr)}}\;.
\end{eqnarray}\end{mathletters}
\par In order for $n_{\mu}$, to be the normal vector field to a 2-D
fast wave-front in curved space requires that the hypersurface orthogonality condition be satisfied,
$n_{[\mu;\nu}n_{\lambda]}=0$: it is a geodesic and  $n_{\mu}=hf,_{\mu}$,
where $f$=constant defines the world-surface of the wave-front and h
is an arbitrary function \citep{lig75}. The geodesic normal vectors
can be described by Carter's equations of geodesic motion
\citep{pun01}. There are 4 constants of motion for a null geodesic, the mass is zero, $m$ is the
angular momentum about the symmetry axis of the black hole, $\omega$ is the
energy of the geodesic and $K^{2}$ is Carter's fourth constant of
motion which represents the relativistic total angular momentum of the
geodesic. By axisymmetry of the wave front and
the light-like velocity of the wave we have $m=0$. Using these
definitions in the hypersurface orthogonality condition implies that
the outgoing (in the stationary frames) fast ffde wave-fronts around a
Kerr black hole are defined by $f$= constant surfaces and $h=1$,
\begin{eqnarray}
&&
f=\pm\int\sqrt{K^{2}-\omega^{2}a^{2}\sin^{2}{\theta}}\,d\theta+\int\frac{\sqrt{\omega^{2}\left(r^{2}+a^{2}\right)-\Delta
  K^{2}}}{\Delta}\,dr-\omega t\;.
\end{eqnarray}The quantities $\omega$ and $K^{2}$ are constants along the
wave surface in order to satisfy the hypersurface orthogonality
condition (Note that there are no outgoing
spherical fast waves near the black hole in the Kerr geometry). In the
proper frame, $n_{\mu}\equiv
(-\mid\mathbf{\bar{n}}\mid,\;\mathbf{\bar{n}})$. Since $m=0$, $\bar{n}^{\phi}=0$. The other components of the wave normal vector
field are found by using (4.1) and (4.2) to evaluate $f,_{\mu}$:
\begin{eqnarray}
&& \bar{n}^{r}= \frac{\gamma}{\rho\sqrt{\Delta}}\left[2Mr\omega+\sqrt{\omega^{2}\left(r^{2}+a^{2}\right)^{2}-\Delta
  K^{2}}\right]\;,\;\bar{n}^{\theta}=\pm\frac{1}{\rho}\sqrt{K^{2}-\omega^{2}a^{2}\sin^{2}{\theta}}\;.
\end{eqnarray}
 \subsection{Causal Stress-Energy Constraints} The equivalence
principle demands that $T^{\mu\nu}$ of the waves is well-behaved in
a freely falling frame, this is the fundamental physical reason why a
black hole can not radiate an \textbf{arbitrary} spectrum and flux of
light waves \citep{can80}. In this subsection, the same result is
demonstrated for ffde fast waves. It is straightforward to evaluate
the structure of a transverse step wave in the \textbf{upstream} proper frame. From the polarization properties in section 2, the
fields transported by the fast wave are determined by
$\mathbf{\bar{n}}$ and $\mathbf{b}_{u}$ in terms of a potential
function, $\bar{E}_{\theta}$,
\begin{eqnarray}
&&
\mathbf{\bar{E}}=\left[-\frac{\bar{n}_{\theta}}{\bar{n}_{r}}\bar{e}_{r}+\bar{e}_{\theta}+\frac{\bar{n}_{\theta}b^{r}}{\bar{n}_{r}b^{\phi}}\bar{e}_{\phi}\right]\bar{E}_{\theta}\,,
\mathbf{\bar{B}}=\left[\frac{\bar{n}_{\theta}^{2}b^{r}}{\bar{n}_{r}\mid\mathbf{\bar{n}}\mid b^{\phi}}\bar{e}_{r}-\frac{\bar{n}_{\theta}b^{r}}{\mid\mathbf{\bar{n}}\mid b^{\phi}}\bar{e}_{\theta}+\frac{\mid\mathbf{\bar{n}}\mid}{\bar{n}_{r}}\bar{e}_{\phi}\right]\bar{E}_{\theta}\;.
\end{eqnarray}
By (4.1), (4.4) and (3.3) if the wave propagates variations,
$\delta\Omega_{_{F}}$ then $\bar{E}_{\theta}$ transported
by the wave in the upstream proper frame is
\begin{eqnarray}
&& \bar{E}_{\theta}=-\frac{\bar{n}^{r}\sqrt{g_{\phi\phi}}
  B^{P}}{\gamma\left[\bar{n}^{r}+v^{r}\mid\mathbf{\bar{n}}\mid\right]c\alpha}\delta\Omega_{_{F}}+O\left(\left[\frac{\bar{n}^{\theta}}{\mid\mathbf{\bar{n}}\mid}\right]^{2}\right)\;, 
\end{eqnarray}where the error terms represent the small changes to
  $B^{P}$ transported by the wave. According to (4.4) and (4.5), outgoing fast
  waves ($\bar{n}^{r}>0$) are blueshifted near the horizon in a proper
  frame: $\bar{B}_{\phi}\approx\bar{E}_{\theta}\sim\delta\Omega_{_{F}}\alpha^{-2}$. By
  contrast for ingoing waves, the fields are well behaved in the proper
  frame $\sim\alpha^{0}$: this is the essence of the ``ingoing wave
  condition'' that is used as a constraint on the background fields
  near the horizon in ffde. The Poynting flux of the outgoing wave has
  two divergent contributions in the upstream proper frame, the linear term,
  $\mathbf{E}\times\mathbf{b}_{u}\sim\alpha^{-2}$ and the dominant, quadratic pure
  wave contribution $T^{\mu\nu}\sim
  n^{\mu}n^{\nu},\,\mathbf{E}\times\mathbf{B}\sim\alpha^{-4}$. The
  same result occurs in a freely falling frame by (3.19) of
  \citet{pun01}. Alternatively stated, the regularity of $T^{\mu\nu}$ in a freely falling
frame near the horizon requires that a globally outgoing fast wave
  near the horizon can only affect changes in
  $\delta\Omega_{_{F}}\sim\alpha_{0}^{2}$, where the subscript ``0'' means to evaluate at the point wave emission (a similar result is true for the
  poloidal current). The exact same scaling is found from detailed
  calculations of outgoing MHD waves near the horizon (see Table 6.1
  of \citet{pun01}). Formally in ffde, the event horizon cannot
  radiate changes in $\Omega_{_{F}}$, or the poloidal current. 
\par This is a significant result in the steady state. By (1.1),
$\mathbf{J}\cdot\mathbf{E}=0$, so there is no transfer of energy from
the plasma to the fields in the magnetosphere. Thus the poloidal current
that supports the Poynting flux cannot be created within the
magnetosphere in the steady state solution (in contrast to MHD). It
must be injected at the boundaries. The only boundaries on the field
aligned current are the event horizon and asymptotic infinity. But,
the horizon cannot change the poloidal current, creating another causality paradox.
\par The simulation evolves through large changes to
$\Omega_{_{F}}$ that emerge from near the horizon. The pathology of
these waves cannot be dismissed by moving the point of emission just
outside the horizon. It arises from physical contradiction of large
amplitude force-free waves emerging from a region that is inertially
dominated - gravity determines the global particle trajectories
irrespective of all external forces \citep{pun01}. One can understand
this inconsistency by considering the effect of one of these waves,
emerging from near the horizon, on the plasma upstream in the proper
frame. By (4.5) the fields transported by the wave diverge like
$\alpha^{-2}$, and such a perturbation drastically alters the direction of $\mathbf{b}$. Thus, there
are large accelerations of the particles as they transition from one
force-free state to another as the wave-front passes: flowing parallel to $\mathbf{b}_{u}$ upstream to
flowing parallel to the perturbed $\mathbf{b}_{d}$, downstream. The
physics of this interaction cannot be described within ffde. These large accelerations imply strong forces as well,
no matter how tenuous the plasma. This information is lost in ffde by setting the mass equal to zero in
MHD. However, full MHD calculations capture this strong force. Even in the most tenuous of plasma states, any MHD fast wave requires very
strong inertial, cross-field (nonforce-free) currents in order to
propagate outwards from near the event horizon (see (6.93) of
\citet{pun01} or \citet{hir93}). These currents translate to a very
strong $F^{\mu\nu}J_{\nu}$ force in the proper frame. \textbf{This
  pathology of ffde is a direct consequence of the momentum equations
  of the plasma near the horizon, the plasma attains an infinite
  inertia in a global sense \citep{pun01} which is the opposite of the
  force-free assumption that assumes plasma inertia is negligible.} 
\section{Conclusion}In this paper two concerns about ffde simulations
were raised. First of all the event
horizon cannot effect V due to the gravitational redshift. Secondly,
if V is modified by fast waves that emerge from just outside the
horizon then they are created by a plasma that cannot support them
self-consistently. The plasma state is set by the gravitational field
(inertially dominated) and outgoing, large amplitude, plasma waves
interact with a real plasma in this region through strong forces in
the proper or freely falling frames. 
\par The Blandford-Znajek solution is the unique ffde steady state
solution since there is no other ffde solution that conserves energy and
angular momentum in the flux tubes: $B^{T}$ and $\Omega_{_{F}}$
are a constant in each flux tube. Any simulation restricted to
ffde either finds no solution or this one. The only ambiguity is the
very minor modifications to the initial poloidal field noted by
\citet{kom01}. If the ffde restriction is relaxed then there
are many MHD solutions and if the MHD restriction is removed then
there are others as well \citep{pun01}. The paradox of the outgoing
``torsional Alfver wave'' described in the
Introduction is an indication that simulations required to seek this
unique ffde solution are unphysical. The calculations of section 4
demonstrate that the simulation evolves by means of unphysical waves
and currents. Furthermore, the results of this paper conflict with the strong
waves that are instantaneously created in the simulation as a consequence of $\mathbf{J}_D$ in the initial state.  
\par Because of their diametrically opposite outgoing wave properties
(very strong inertial forces versus force-free) near the
horizon, the magnetically dominated limit of a full MHD simulation should differ substantially from this ffde treatment. The results of
the MHD simulations of \citet{koi02}, \citet{cak00}, \citet{sem01} and the
time stationary solution in Chapter 9 of \citet{pun01} indicate that this is the case. This fact is
apparent in all of the above in which relativistic inertia
imparted to the plasma by the gravitational field dominates the
dynamics in the ergosphere, regardless of the degree of inertial
dominance imposed in the initial state. The interaction drives very
strong cross-field currents (note that
these currents cannot be dismissed as transients in \citet{pun01} in
which they are eternal and \citet{sem01} in which they are persistent
for long simulations and the driving force never goes away). Not coincidentally, this is precisely the region in which
unphysical waves emerge in the ffde simulation.
\par In summary, the force-free assumption does not apply to physics
near the event horizon. Even if an initial MHD state is chosen
by hand to be force-free, the pathology of force-free physics near the
horizon becomes evident when the outgoing MHD perturbations of the
initial state (such as the waves discussed above) are analyzed. 

\end{document}